\documentclass[conference,comsoc]{IEEEtran}
\usepackage[T1]{fontenc}% optional T1 font encoding
\usepackage{amsmath}
\usepackage[cmintegrals]{newtxmath}
\usepackage{url}
\usepackage{soul,color,graphicx,float,epstopdf}
\usepackage{tablefootnote,textcomp,gensymb}
\usepackage{eurosym,booktabs,multirow}
\usepackage{tabularx} % Add this package
\usepackage{amsfonts} 
\usepackage{algorithmic}
\usepackage{array}
\usepackage{stfloats}
\usepackage{float}
\usepackage{mathtools}
\usepackage{graphicx}
\usepackage{pdfpages}
\usepackage{comment}
\usepackage{balance}
\usepackage{xcolor}
\usepackage{comment,enumitem}
\usepackage[normalem]{ulem}
\usepackage[implicit=false]{hyperref}
\usepackage{subfigure}
\definecolor{Orange}{rgb}{1,0.5,0}
\newcommand{\todo}[1]{\textsf{\textbf{\textcolor{Orange}{[#1]}}}}

\begin{document}
\title{Flying Base Stations for Offshore Wind Farm Monitoring and Control: Holistic Performance Evaluation and Optimization}

\author{\IEEEauthorblockN{
Xinyi Lin, Peizheng Li, Adnan Aijaz
}\\ 
\IEEEauthorblockA{
Bristol Research and Innovation Laboratory, Toshiba Europe Ltd., U.K.\\
Email: {\{xinyi.lin, peizheng.li, adnan.aijaz\}@toshiba-bril.com}}
}

\maketitle

\begin{abstract}
Ensuring reliable and low-latency communication in offshore wind farms is critical for efficient monitoring and control, yet remains challenging due to the harsh environment and lack of infrastructure. This paper investigates a flying base station (FBS) approach for wide-area monitoring and control in the UK Hornsea offshore wind farm project. By leveraging mobile, flexible FBS platforms in the remote and harsh offshore environment, the proposed system offers real-time connectivity for turbines  without the need for deploying permanent infrastructure at the sea. We develop a detailed and practical end-to-end latency model accounting for five key factors: flight duration, connection establishment, turbine state information upload, computational delay, and control transmission, to provide a holistic perspective often missing in prior studies. Furthermore, we combine trajectory planning, beamforming, and resource allocation into a multi-objective optimization framework for the overall latency minimization, specifically designed for large-scale offshore wind farm deployments. Simulation results verify the effectiveness of our proposed method in minimizing latency and enhancing efficiency in FBS-assisted offshore monitoring across various power levels, while consistently outperforming baseline designs.
\end{abstract}


\section{Introduction}
\label{sec:introduction}
The UK Hornsea Project is one of the world's largest offshore wind farm projects, located in the North Sea off the east coast of England, aiming to boost renewable energy and enhance UK energy security through large-scale offshore wind power. The project is divided into multiple phases, including Hornsea One~\cite{orsted_hornsea1}, Two~\cite{orsted_hornsea2}, and Three~\cite{orsted_hornsea3}, with a total planned capacity exceeding 5GW. Hornsea One (1.2GW, operational since 2020) and Hornsea Two (1.3GW) have significantly expanded the UK’s offshore wind capacity. Expected to be completed by 2027, Hornsea Three will incorporate cutting-edge turbines, smart grid integration, and environmentally conscious design with net capacity of 2.9GW. 
%Hornsea One, which started operating in 2020, has a capacity of 1.2 GW, making it the largest offshore wind farm at the time. Hornsea Two, with a 1.3 GW capacity, further expanded the project’s contribution to renewable energy. Project Three builds on the success of its predecessors, Hornsea One and Hornsea Two, which are already operational and among the largest offshore wind farms globally. Hornsea Three will feature state-of-the-art wind turbines, advanced grid integration technologies, and innovative solutions to minimize environmental impact. It is expected to play a critical role in reducing carbon emissions, enhancing energy security, and supporting the UK's transition to a net-zero economy by 2027. 
When fully developed, the Hornsea Project is expected to power millions of UK homes, significantly advancing the country’s commitment to clean energy and reducing carbon emissions. 

The Hornsea wind farm, while a remarkable achievement in renewable energy, still faces significant challenges related to communication, monitoring, and control due to its remote location, harsh environment, and large scale. Specifically, the wind farms are located approximately 120km offshore, making it difficult to establish reliable, low-latency communication links for real-time monitoring and control. In addition, offshore wind farms require real-time monitoring of turbine performance, environmental conditions, and grid stability, but traditional systems may struggle to provide high-resolution, low-latency data. Moreover, installing and maintaining permanent communication infrastructure for offshore wind farms is costly and may not be scalable for future expansions.

Flying Base Stations (FBS)~\cite{8011325}, also referred to as UAV-mounted base stations~\cite{9314048}, offer an innovative solution to these challenges by providing mobile, flexible, and cost-effective communication and monitoring capabilities. By integrating FBS with wind farms' operations, operators can improve real-time monitoring, enhance grid stability, and ensure the reliable and efficient operation of this critical renewable energy asset.

Some current research directions and real-world applications of FBS highlight their potential and ongoing investigations. For example, the authors in~\cite{8796414} exploited how to optimally deploy a minimized number of UAVs to achieve on-demand coverage for ground users. Moreover, a novel communication coverage improvement strategy was proposed in~\cite{9515712}, which jointly selects the optimal clusters of an arbitrary number of the users served at the same time-frequency resources by means of non-orthogonal multiple access (NOMA), allocates the optimal transmission power to each user, and determines the position of the FBS. The authors in~\cite{8247211} considered a multi-UAV enabled wireless communication system. To achieve fair performance among users, the minimum throughput over all ground users in the downlink communication was maximized by optimizing the multiuser communication scheduling and association jointly with the UAV's trajectory and power control. However, existing studies on FBS have primarily focused on urban or terrestrial environments, often optimizing coverage, throughput, or energy efficiency for ground users. These approaches generally lack applicability to complex, large-scale offshore environments such as the discussed Hornsea wind farm, where challenges like long-distance communication, harsh weather conditions, and distributed infrastructure demand specialized solutions. The dynamic nature of operations and the need for real-time monitoring are often overlooked as well, which are key factors in the wind farm scenario. 

To address the challenges of large-scale offshore wind farm monitoring and control, we propose a comprehensive FBS design framework specifically tailored for Hornsea Wind Farm. Owing to the extensive coverage area of such projects, Hornsea Two, for example, spans 462 square kilometers (178 square miles), which is 31 times the size of Lake Windermere, the deployment of FBSs poses unique challenges. Given that FBSs are typically battery-powered with limited energy capacity, failing to incorporate energy-aware designs such as trajectory planning may result in premature battery depletion, rendering the system unable to support continuous operation. Therefore, power consumption and latency emerge as critical factors that must be carefully considered in this context. To this end, we develop an FBS design strategy that jointly optimizes trajectory planning, communication connection, uplink and downlink beamforming, power allocation, and latency minimization, while accounting for practical constraints such as energy consumption limits and signal-to-noise ratio (SNR) requirements. By effectively addressing these key challenges, our framework enhances real-time monitoring and control, significantly improving efficiency, reliability, and responsiveness of the wind farm system. The main contributions of this work are summarized as follows.
\begin{itemize}
    \item This paper proposes the use of FBS to improve real-time monitoring and control of offshore wind farms, particularly focusing on the UK Hornsea offshore wind farm. It highlights the challenges in communication due to long distances, harsh weather conditions, and the lack of permanent infrastructure, and presents FBS as a mobile, flexible, and cost-effective solution.
    \item Based on the offshore wind farms scenario, a detailed latency model is developed by considering five key components: flight duration, connection establishment, turbine state information upload, computational time, and payload control signal transmission. This provides a holistic view of delays in FBS-assisted wind farm monitoring, which is often overlooked in previous research.
    \item We apply a combination of trajectory planning, optimized communication, and resource allocation techniques to minimize latency and improve efficiency in FBS-assisted offshore wind farm monitoring. This multi-objective optimization framework is tailored for large-scale offshore wind farms. Simulation results verify the effectiveness of our proposed strategy.
\end{itemize}

\section{System Model}
 \begin{figure}[t]
    \centering    \includegraphics[width=1\linewidth]{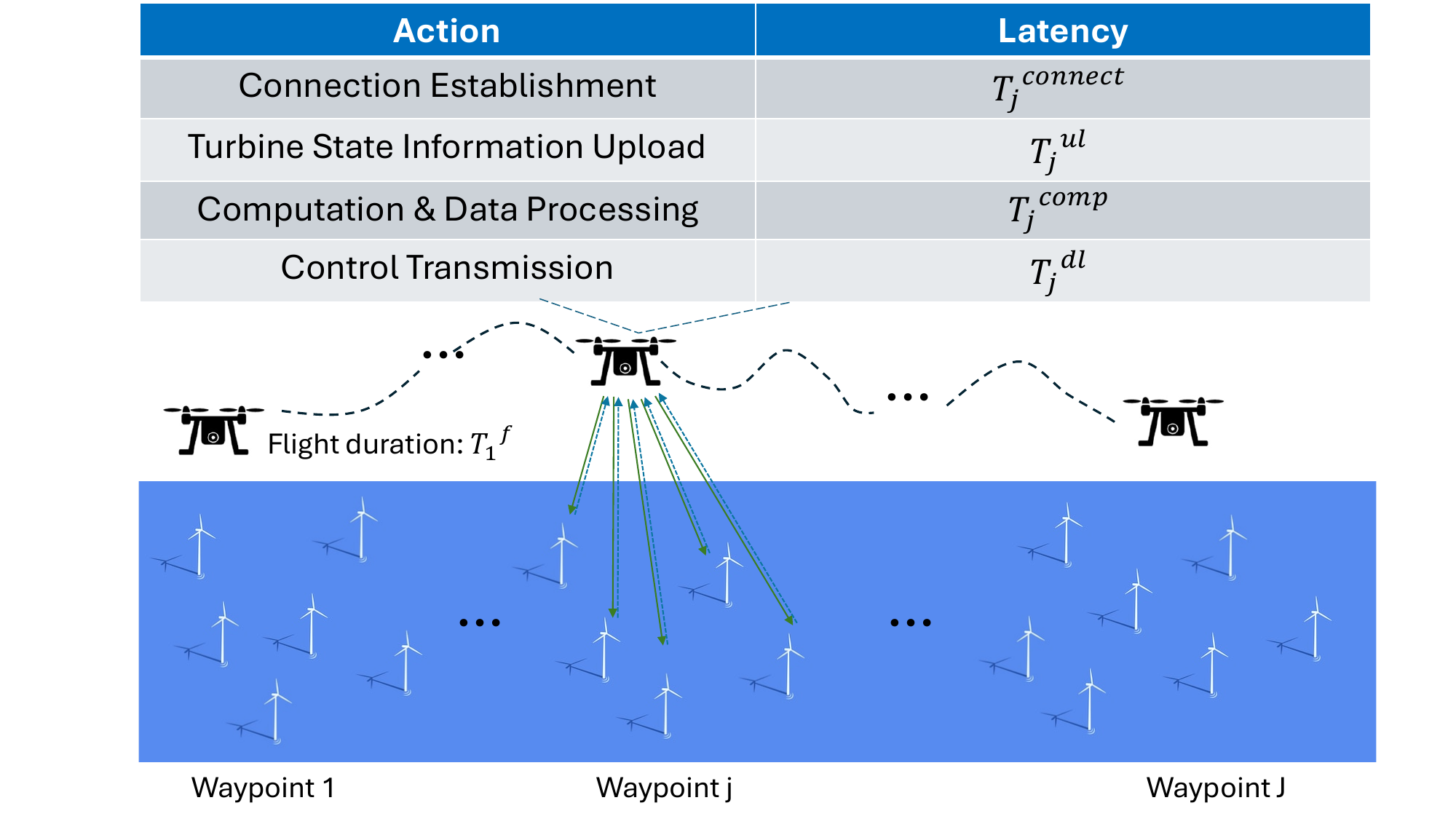}
    \caption{System model of the FBS assisted wide area monitoring and control. }
    \label{systemmodel}
\end{figure}

Fig. \ref{systemmodel} depicts the system model where a FBS flies across multiple waypoints in the wind farm for turbine monitoring and control. The FBS collects turbine state information (e.g., wind speed, rotor speed, pitch angle) and transmits optimized control signals to adjust the turbine operation. We first define the system latency as the sum of the following five components:
\begin{itemize}
    \item Flight Duration: The time required for the FBS to travel from the source to the destination.
\item Connection Establishment: The time needed for the FBS to establish a connection with the turbines at each waypoint.
\item Turbine State Information Upload: The time taken to transmit turbine state data to the FBS.
\item Computation Time: The time the FBS spends performing edge AI inference to generate real-time control actions.
\item Control Transmission: The time required to transmit optimized control signals to adjust turbine operations.
 \end{itemize}

 To model the latencies, we assume that there are a total of $J$ waypoints in the set $\mathcal{J}$ and $K$ turbines in the set $\mathcal{K}$. Each turbine is assigned to communicate with the FBS at a specific waypoint for monitoring and control purposes. Particularly, when the FBS is located at the $j$-th waypoint, a number of $K_j$ turbines send data, such as operational status, sensor readings, and other monitoring information, up to the FBS. The FBS acts as a base station, which is implemented with $N$ elements, receiving this data and performing the necessary monitoring and control operations. The set of the $K_j$ turbines is denoted as $\mathcal{K}_j$. Each turbine contains $M$ antenna elements. The uplink and downlink channel between the $k$-th turbine and the FBS is denoted as $\boldsymbol{H}_k\in\mathbb{C}^{N\times M}$ and $\boldsymbol{G}_k\in\mathbb{C}^{M\times N}$, respectively. We assume that the carrier frequency is 3.85 GHz.
The channels are modeled as \cite{10364735}
  \begin{equation}
     \boldsymbol{H}_k = \text{PL}_{\text{LoS},k}(\sqrt{\frac{\varepsilon}{\varepsilon+1}}\boldsymbol{a}(\boldsymbol{\zeta}_{k})\boldsymbol{a}^H(\boldsymbol{\zeta}_{BS})+\sqrt{\frac{1}{\varepsilon+1}}\overline{\boldsymbol{H}_k}),
 \end{equation}
 \begin{equation}
     \boldsymbol{G}_k = \text{PL}_{\text{LoS},k}(\sqrt{\frac{\varepsilon}{\varepsilon+1}}\boldsymbol{a}(\boldsymbol{\zeta}_{BS})\boldsymbol{a}^H(\boldsymbol{\zeta}_{k})+\sqrt{\frac{1}{\varepsilon+1}}\overline{\boldsymbol{G}_k}),
 \end{equation}
Specifically, $\text{PL}_{\text{LoS},k}$ denotes the corresponding path loss, which involves the rainfall attenuation and atmospheric gas attenuation. $\varepsilon$ is the Rician factor. $\overline{\boldsymbol{H}_k}$ and $\overline{\boldsymbol{G}_k}$ denote the non-line-of-sight (NLoS) components of channels, each element of which follows $\mathcal{CN}(0,1)$. In addition, $\boldsymbol{a}(\boldsymbol{\zeta})$ denotes the steering vector as a function of the spatial information $\boldsymbol{\zeta}$, which is represented as
\begin{equation}
    \boldsymbol{a}(\boldsymbol{\zeta})=e^{j\frac{2\pi d}{\lambda}\boldsymbol{\zeta}}.
\end{equation}
Note that $\boldsymbol{\zeta}$ can follow either the uniform linear array (ULA) or uniform rectangular array (URA) configuration. Moreover, $\boldsymbol{\zeta}_{k}$ is the function of angles of departure (AoD) from the $k$-th turbine towards the FBS, and $\boldsymbol{\zeta}_{BS}$ denotes the function of angles of arrival (AoA) at the FBS. Specifically, we consider the URA configuration for both the FBS and turbines in our proposed system, then $\boldsymbol{\zeta}_{k}$ and $\boldsymbol{\zeta}_{BS}$ can be generalized as $\boldsymbol{\zeta} = f(\varphi,\theta)$, where $\varphi$ and $\theta$ refer to the azimuth and elevation angle, respectively. Assuming that for each turbine, there are $M_x$ elements along the $X$-axis and $M_y$ elements along the $Y$-axis, then we have $M=M_x M_y$. Similarly, we have $N=N_xN_y$ at FBS. In addition, we denote the azimuth angle $\varphi_{k,m}$ and the elevation angle $\theta_{k,m}$ as the departing angles from the $m$-th element in the $k$-th turbine, and $\varphi_{BS,n}$ and $\theta_{BS,n}$ as the azimuth and elevation angles towards the $n$-th element in the FBS. Note that $m$ has a relationship with $m_x$ and $m_y$ as $m = (m_x-1) M_y+m_y$, where $m_x\in\{1, \cdots, M_x\}$, $m_y\in\{1,\cdots, M_y\}$. The similar relationship can be derived for $n$, $n_x$, and $n_y$ as well. The angle information $\zeta_{k,m}$ and $\zeta_{BS,n}$ could be represented as
{\begin{equation}
    \zeta_{k,m} = (m_x-1) \cos\theta_{k,m}\cos\varphi_{k,m}+(m_y-1)\cos\theta_{k,m}\sin\varphi_{k,m},
\end{equation}
\begin{equation}
    \zeta_{BS,n} = (n_x-1) \cos\theta_{BS,n}\cos\varphi_{BS,n}+(n_y-1)\cos\theta_{BS,n}\sin\varphi_{BS,n}.
    \label{Omega_out}
\end{equation}}
Sequentially, $\boldsymbol{\zeta}_{k}$ and $\boldsymbol{\zeta}_{BS}$ are denoted as $\boldsymbol{\zeta}_{k}=[\zeta_{k,1},\cdots,\zeta_{k,m},\cdots,\zeta_{k,M}]^T,\forall k\in \mathcal{K}_j$ and $\boldsymbol{\zeta}_{BS}=[\zeta_{BS,1},\cdots,\zeta_{BS,n},\cdots,\zeta_{BS,N}]^T$.
Let $\boldsymbol{W}_{ul,j}=[\boldsymbol{w}_{ul,1},\boldsymbol{w}_{ul,2},...,\boldsymbol{w}_{ul,K_j}]$ and $\boldsymbol{V}_{ul,j}=[\boldsymbol{v}_{ul,1},\boldsymbol{v}_{ul,2},...,\boldsymbol{v}_{ul,K_j}]$ denote the beamforming matrix of the FBS and the turbines in the uplink system, respectively.
The SINR for the $k$-th turbine could be expressed as
\begin{equation}
\text{SINR}_{ul,k}=\frac{\lvert\boldsymbol{w}_{ul,k}^H\boldsymbol{H}_{k}\boldsymbol{v}_{ul,k}\rvert^2 P_k}{\sum_{l\neq k}\lvert\boldsymbol{w}_{ul,k}^H\boldsymbol{H}_l\boldsymbol{v}_{ul,l}\rvert^2 P_l+\sigma^2},
\end{equation}
where $\sigma^2$ represents the noise power. $P_k$ denotes the transmit power from the $k$-th turbine. Similarly, the SINR for the $k$-th turbine in the downlink system can be expressed as
\begin{equation}
    \text{SINR}_{dl,k} = \frac{\lvert\boldsymbol{v}_{dl,k}^H\boldsymbol{G}_{k}\boldsymbol{w}_{dl,k}\rvert^2 P_{dl,k}}{\sum_{l\neq k}\lvert\boldsymbol{v}_{dl,k}^H\boldsymbol{G}_k\boldsymbol{w}_{dl,l}\rvert^2 P_{dl,l}+\sigma^2},
\end{equation}
where $\boldsymbol{W}_{dl,j}=[\boldsymbol{w}_{dl,1},\boldsymbol{w}_{dl,2},...,\boldsymbol{w}_{dl,K_j}]$ and $\boldsymbol{V}_{dl,j}=[\boldsymbol{v}_{dl,1},\boldsymbol{v}_{dl,2},...,\boldsymbol{v}_{dl,K_j}]$ denote the beamforming matrix of the FBS and the turbines in the downlink system, respectively. $P_{dl,k}$ denotes the power allocated for the downlink communication between FBS and the $k$-th turbine. Then, the uplink and downlink rate of the $k$-th turbine is represented as
\begin{equation}
    R_{\{ul,dl\},k}=\log_2(1+\text{SINR}_{\{ul,dl\},k}).
\end{equation}
We assume that the $k$-th turbine intends to send data and receive control information with the size of $D_{ul,k}$ and $D_{dl,k}$, respectively. %For simplicity, we assume that all turbines transmit information with the same size, i.e., $D_{\{ul,dl\},k} = D_{\{ul,dl\}}, \forall k\in \mathcal{K}$ \footnote{The assumption does not affect the generality of the problem and can be readily extended to more general cases.}. 
The uplink and downlink transmission latency at the $j$-th waypoint can be obtained as
\begin{equation}
    T_j^{\{ul,dl\}}=\max\{\frac{D_{\{ul,dl\},k}}{R_{\{ul,dl\},k}}, \forall k\in \mathcal{K}_j\}.
\end{equation}

Furthermore, we define the computational processing delay of the FBS at $j$-th waypoint as $T_j^{comp}$, which could be expressed as the sum of the delay for $K_j$ turbines:
\begin{equation}
    T_j^{comp}=\sum_{k=1}^{K_j}T_{j,k}^{comp}.
\end{equation}
Particularly, the processing delay of the information from the $k$-th turbine can be expressed as
\begin{equation}
    T_{j,k}^{comp}=\frac{C_k}{f_k},
\end{equation}
where $C_k=\mu D_{ul}$ denotes the computation workload for turbine $k$. $\mu$ represents the computational intensity (cycles per bit). Additionally, $f_k$ is the allocated CPU frequency (cycles per second) for turbine $k$, which can be interpreted as \cite{888701}
\begin{equation}
    f_k=\left(\frac{P_{comp,k}}{\varsigma}\right)^{\frac{1}{3}},
\end{equation}
 where $\varsigma$ is a hardware-dependent coefficient that reflects the energy efficiency of the processor, and $P_{comp,k}$ is the power allocated for processing the information from turbine $k$. Combining the power allocated for downlink transmission and computation processing, we derive the power allocation matrix of the FBS at waypoint $j$ as $\boldsymbol{P}_j=\begin{bmatrix}
P_{dl,1}, & \cdots, & P_{dl,K_j}\\
P_{comp,1}, & \cdots, & P_{comp,K_j}
\end{bmatrix}\in \mathbb{C}^{2\times K_j}, \forall j\in \mathcal{J}$.
%Sequentially, the sum rate of turbines covered by the FBS can be represented by
%\begin{equation}
%R=\sum_{l=1}^{L}\log_2(1+\text{SINR}_l).
%\end{equation}
%It is known that $f(\boldsymbol{\zeta}_l)$ will be maximized when $\boldsymbol{b}=\boldsymbol{a}_{BS}^H$. 

We further define the flight duration as $T^f$, and the connection establishment time at each waypoint as $T_j^{connect}$. The overall latency can be expressed as
\begin{equation}
    T=T^f+\sum_{j=1}^J \bigl(T_j^{connect}+T_j^{ul}+T_j^{comp}+T_j^{dl}\bigr).
\end{equation}
A latency minimization problem is then formulated via jointly optimizing the power allocation at the FBS and the beamforming at turbines and the FBS during both uplink and downlink transmission, which is represented as
\begin{subequations}
\begin{align}
\begin{split}
\mathbb { P }: & \min \limits_{\boldsymbol{W}_{\{ul,dl\},j},\boldsymbol{V}_{\{ul,dl\},j}, \boldsymbol{P}_j, \forall j\in \mathcal{J}} \quad T
\label{obj}
\end{split}\\
\begin{split}
 & \qquad \quad \quad \text { s.t. }  \ \text{\text{SINR}}_{\{ul,dl\},k}\geq\text{\text{SINR}}_{th},  \forall k\in \mathcal{K}, 
 \label{19b}
\end{split}\\
\begin{split}
 & \qquad \qquad \qquad \ \lVert\boldsymbol{w}_{\{ul,dl\},k}\rVert_2\leq1, \forall k\in \mathcal{K}_j, 
 j\in\mathcal{J},
 \label{19c}
\end{split}\\
\begin{split}
 & \qquad \qquad \qquad \ \lVert\boldsymbol{v}_{\{ul,dl\},k}\rVert_2\leq1, \forall k\in \mathcal{K}_j, 
 j\in\mathcal{J},
 \label{19d}
 \end{split}\\
 \begin{split}
 & \qquad \qquad \quad \quad \sum_{k=1}^{K_j}\bigl(P_{dl,k}+P_{comp,k}\bigr)\leq P_{j}, \forall j\in \mathcal{J}.
 \label{19e}
\end{split}
\end{align}
\end{subequations}
Particularly, \eqref{19b} gives the SINR constraint for each turbine; \eqref{19c} and \eqref{19d} denote the unit norm constraint at the FBS and the turbines, respectively. \eqref{19e} refers to the power constraint at the FBS. In the following, we propose the design and optimization solution for the latency minimization problem.

\section{Proposed Latency Minimization Solution for FBS and Turbines Communication}

\begin{figure}[!t] 
    \centering  
    \subfigure[View location of turbines in Google Earth.]{
                    \includegraphics[width=0.9\linewidth]{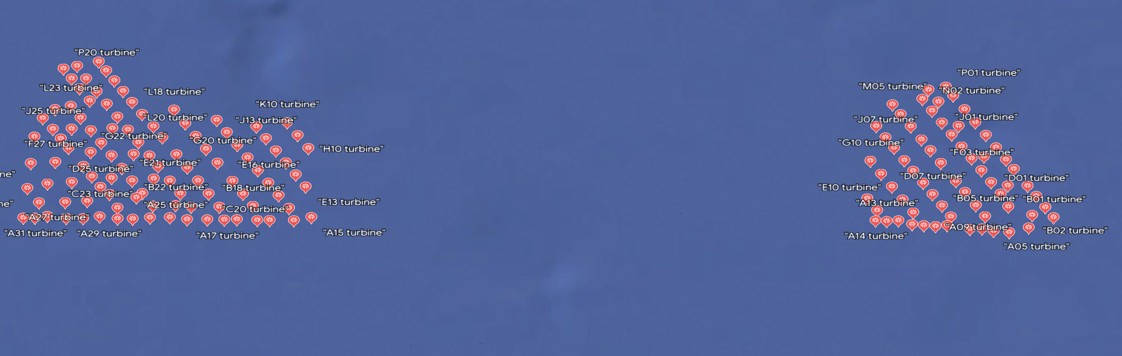}
                }
        \subfigure[UAV path planning based on deployed waypoints and the Dijkstra algorithm.]{
                    \includegraphics[width=0.9\linewidth]{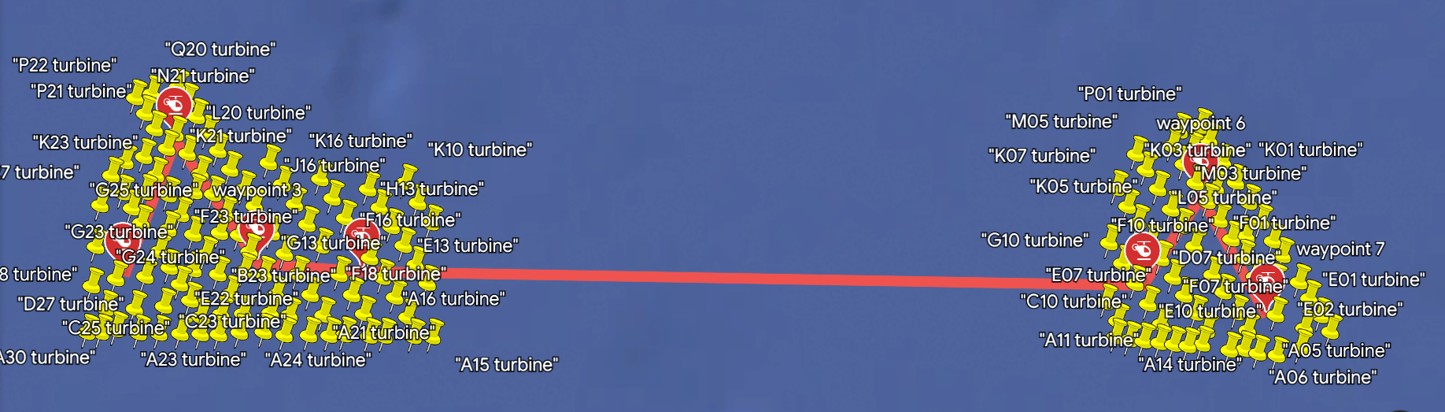}}
    \caption{An overview of the UK Hornsea Offshore Wind Farms project.}
	\label{fig1}
\end{figure}

\begin{figure}
    \centering    \includegraphics[width=0.96\linewidth]{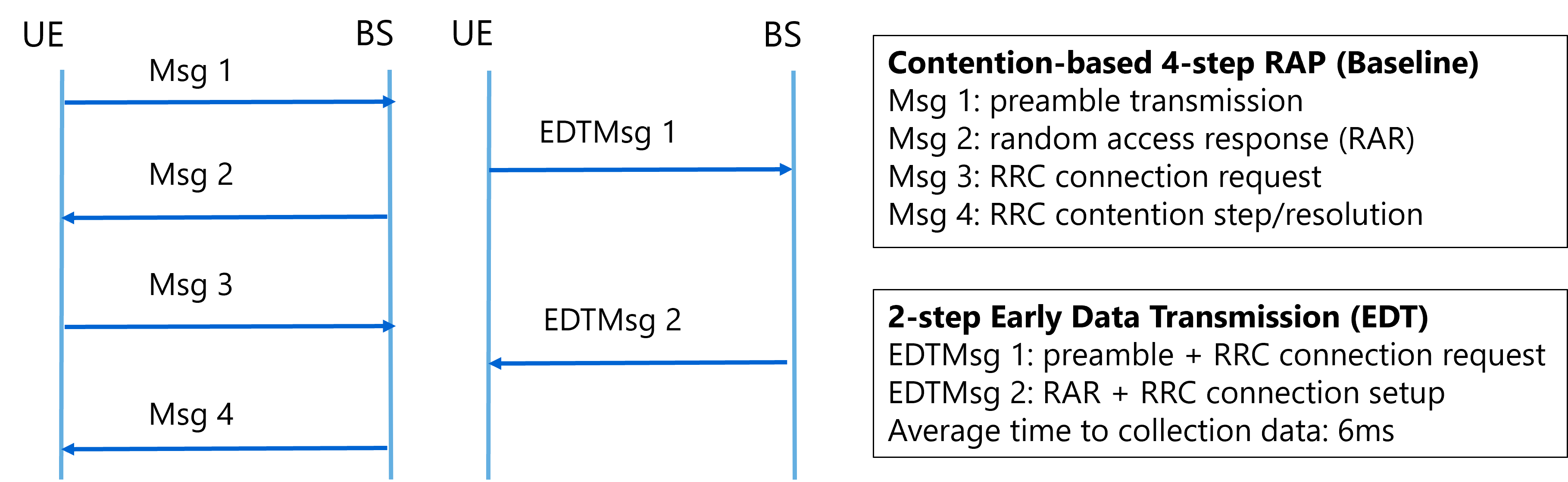}
    \caption{The comparison between the contention-based 4-step RAP and the proposed 2-step EDT scheme.}
    \label{fig2}
\end{figure}

 Fig. \ref{fig1} depicts an overview of the Hornsea Offshore Wind Farms project Two in the UK. In order to realize the minimized flight duration, we first design the path of the UAV based on the deployed waypoints and implement the Dijkstra algorithm \cite{8073641} for the shortest distance. In addition, a 2-step random access with early data transmission (EDT) framework is proposed for the data collection procedure, which effectively decreases FBS's working time. Fig. \ref{fig2} demonstrates the comparison between the contention-based 4-step random access procedure (RAP) and the proposed 2-step EDT scheme \cite{9268113}. Specifically, the contention-based 4-step RAP involves separate steps for preamble transmission, random access response, RRC connection request, and contention resolution, whereas the 2-step EDT scheme combines preamble and RRC connection request in one step and RAR and RRC connection setup in another, significantly reducing signaling overhead and time. 
In the following, we further develop the optimization for the uplink, computation processing, and the downlink phases. 

\subsection{Uplink latency design}
First, the uplink latency can be considered as a function of the beamforming matrix $\boldsymbol{W}_{ul,j}$ and $\boldsymbol{V}_{ul,j}, \forall j\in \mathcal{J}$. Then, the sub-optimal problem can be interpreted as
\begin{subequations}
\begin{align}
\begin{split}
\mathbb { P }1: & \min \limits_{\boldsymbol{W}_{ul},\boldsymbol{V}_{ul}} \quad \sum_{j=1}^{J}T_j^{ul}
\end{split}\\
\begin{split}
 & \quad \text { s.t. }  \quad\text{\text{SINR}}_{ul,k}\geq\text{\text{SINR}}_{th},  \forall k\in \mathcal{K}, 
\end{split}\\
\begin{split}
 & \qquad \qquad \lVert\boldsymbol{w}_{ul,k}\rVert_2\leq1, \forall k\in \mathcal{K}_j, 
 j\in\mathcal{J},
\end{split}\\
\begin{split}
 & \qquad \qquad \lVert\boldsymbol{v}_{ul,k}\rVert_2\leq 1, \forall k\in \mathcal{K}_j, 
 j\in\mathcal{J}.
 \end{split}
\end{align}
\end{subequations}
As the beamforming at each waypoint is independent of the others, we focus on the beamforming design at the $j$-th waypoint in particular, while the beamforming at other waypoints can be similarly derived. We first reformulate the min-max problem as a convex-concave optimization using an auxiliary variable $\tau_j$:
\begin{subequations}
\begin{align}
\begin{split}
\mathbb { P }1.1: & \min \limits_{\boldsymbol{W}_{ul,j},\boldsymbol{V}_{ul,j}} \quad \frac{1}{\tau_j}
\label{obj1}
\end{split}\\
\begin{split}
 & \quad \text { s.t. }  \quad T_{j,k}^{ul}\leq \frac{1}{\tau_j}, \forall k\in \mathcal{K}_j,
 \label{20a}
 \end{split}\\
\begin{split}
 & \qquad \qquad \text{\text{SINR}}_{ul,k}\geq\text{\text{SINR}}_{th},  \forall k\in \mathcal{K}_j, 
 \label{20b}
\end{split}\\
\begin{split}
 & \qquad \qquad \lVert\boldsymbol{w}_{ul,k}\rVert_2\leq1, \forall k\in \mathcal{K}_j,
 \label{20c}
\end{split}\\
\begin{split}
 & \qquad \qquad \lVert\boldsymbol{v}_{ul,k}\rVert_2\leq 1, \forall k\in \mathcal{K}_j.
 \label{20d}
 \end{split}
\end{align}
\end{subequations}
As the variables are coupled, we apply the alternating optimization (AO) algorithm to update $\boldsymbol{W}_{ul,j}$ and $\boldsymbol{V}_{ul,j}$ iteratively. We first fix $\boldsymbol{W}_{ul,j}$ and optimize $\boldsymbol{V}_{ul,j}$. Particularly, \eqref{20a} can be interpreted as
\begin{equation}   \frac{\lvert\boldsymbol{w}_{ul,k}^H\boldsymbol{H}_{k}\boldsymbol{v}_{ul,k}\rvert^2 P_k}{\sum_{l\neq k}\lvert\boldsymbol{w}_{ul,k}^H\boldsymbol{H}_l\boldsymbol{v}_{ul,l}\rvert^2 P_l+\sigma^2}\geq 2^{D_{ul,k}\tau_j}-1.
\label{trans1}
\end{equation}
%We define $\rho_k=\max\biggl(2^{\frac{D_{ul,k}}{\tau_j}}-1, \text{SINR}_{th}\biggr)$, and $I_k=\sum_{l\neq k}\lvert\boldsymbol{w}_{ul,k}^H\boldsymbol{H}_l\boldsymbol{v}_{ul,l}\rvert^2 P_l+\sigma^2$. 
To deal with the fractional programming (FP) problem \cite{10746626}, we approximate the SINR expression as
\begin{multline}
    \text{SINR}_k\approx 2\Re\biggl\{\alpha_k^{(t)}\sqrt{P_k}\lvert\boldsymbol{w}_{ul,k}^H\boldsymbol{H}_{k}\boldsymbol{v}_{ul,k}\rvert\biggr\}-\\(\alpha_k^{(t)})^2\biggl(\sum_{l\neq k}\lvert\boldsymbol{w}_{ul,k}^H\boldsymbol{H}_l\boldsymbol{v}_{ul,l}\rvert^2 P_l+\sigma^2\biggr),
    \label{fp1}
\end{multline}
where
\begin{equation}
    \alpha_k^{(t)} = \frac{\Re\biggl\{\sqrt{P_k}\lvert\boldsymbol{w}_{ul,k}^H\boldsymbol{H}_{k}\boldsymbol{v}_{ul,k}\rvert\biggr\}}{\sum_{l\neq k}\lvert\boldsymbol{w}_{ul,k}^H\boldsymbol{H}_l\boldsymbol{v}_{ul,l}\rvert^2 P_l+\sigma^2}.
\end{equation}
Substituting \eqref{fp1} into \eqref{trans1}, the convexity of this constraint is ensured.
Furthermore, to address the non-convexity of the SINR threshold constraint, we reformulate it using first-order Taylor expansion, which is approximated as:

\begin{multline}
    |\mathbf{w}_{ul,k}^H \mathbf{h}_k \mathbf{v}^{(t)}_{ul,k}|^2+2\Re\biggl\{(\mathbf{v}^{(t)}_{ul,k})^H\mathbf{G}_k(\mathbf{v}_{ul,k}-\mathbf{v}^{(t)}_{ul,k})\biggr\} \geq \\ \text{SINR}_{th}
   \biggl(\sum_{l\neq k}\lvert\boldsymbol{w}_{ul,k}^H\boldsymbol{H}_l\boldsymbol{v}_{ul,l}\rvert^2 P_l+\sigma^2\biggr),
\end{multline}
where $\mathbf{G}_k=(\mathbf{w}_{ul,k}^H \mathbf{h}_k)^H(\mathbf{w}_{ul,k}^H \mathbf{h}_k)$.

The convex optimization problem with respect to $\boldsymbol{V}_{ul,j}$ is therefore formulated, which could be easily solved by the CVX toolbox. The optimization of $\boldsymbol{W}_{ul,j}$ at the $t$-th can be similarly derived. By iteratively updating $\boldsymbol{V}_{ul,j}$ and $\boldsymbol{W}_{ul,j}$ until convergence, the minimized uplink transmission latency can be obtained.

\subsection{Computation processing and downlink latency design}
As the power of the FBS needs to be optimally allocated to the computing process and downlink transmission, the joint computation processing and downlink latency minimization design is proposed. We focus on the latency minimization at the $j$-th waypoint, then the problem can be interpreted as:
\begin{subequations}
\begin{align}
\begin{split}
\mathbb { P }2: & \min \limits_{\boldsymbol{W}_{dl,j},\boldsymbol{V}_{dl,j}, \boldsymbol{P}_j} \quad T_j^{comp}+T_j^{dl}
\label{obj2}
\end{split}\\
\begin{split}
 &\ \ \qquad \text { s.t. }  \quad\ \text{\text{SINR}}_{dl,k}\geq\text{\text{SINR}}_{th},  \forall k\in \mathcal{K}_j, 
 \label{21b}
\end{split}\\
\begin{split}
 & \qquad \qquad \qquad \lVert\boldsymbol{w}_{dl,k}\rVert_2\leq1, \forall k\in \mathcal{K}_j,
 \label{21c}
\end{split}\\
\begin{split}
 & \qquad \qquad \qquad \lVert\boldsymbol{v}_{dl,k}\rVert_2\leq 1, \forall k\in \mathcal{K}_j,
 \label{21d}
 \end{split}\\
 \begin{split}
 & \qquad \qquad \quad \quad\sum_{k=1}^{K_j}\bigl(P_{dl,k}+P_{comp,k}\bigr)\leq P_{j}.
 \label{21e}
\end{split}
\end{align}
\end{subequations}
We derive the AO algorithm for the iterative update of the variables. As the beamforming optimization with respect to $\boldsymbol{W}_{dl,j}$ and $\boldsymbol{V}_{dl,j}$ is similar to the optimization of $\boldsymbol{W}_{ul,j}$ and $\boldsymbol{V}_{ul,j}$, we omit the derivation here due to the page limits. For the power allocation design at waypoint $j$, we first introduce an auxiliary variable $\gamma_j$ and expand the objective function as
\begin{equation}
    \mathcal{O}(\boldsymbol{P}_j)=\sum_{k=1}^{K_j}\mu D_{ul,k}\varsigma^{\frac{1}{3}}{P_{comp,k}}^{-\frac{1}{3}}+\frac{1}{\gamma_j},
\end{equation}
with
\begin{equation}   \frac{\lvert\boldsymbol{v}_{dl,k}^H\boldsymbol{G}_{k}\boldsymbol{w}_{dl,k}\rvert^2 P_{dl,k}}{\sum_{l\neq k}\lvert\boldsymbol{v}_{dl,k}^H\boldsymbol{G}_k\boldsymbol{w}_{dl,l}\rvert^2 P_{dl,l}+\sigma^2}\geq 2^{D_{dl,k}\gamma_j}-1.
\label{P_fp}
\end{equation}
The objective function is represented in the convex form. However, \eqref{P_fp} contains the coupled variables $\boldsymbol{P}_j$ and $\gamma_j$, making it non-convex. To address the non-convexity, we rewrite the SINR term as
 \begin{multline}
    \text{SINR}_k\approx 2\Re\biggl\{\beta_k^{(t)}\sqrt{P_{dl,k}}\lvert\boldsymbol{v}_{dl,k}^H\boldsymbol{G}_{k}\boldsymbol{w}_{dl,k}\rvert\biggr\}-\\(\beta_k^{(t)})^2\biggl(\sum_{l\neq k}\lvert\boldsymbol{v}_{dl,k}^H\boldsymbol{G}_l\boldsymbol{w}_{dl,l}\rvert^2 P_{dl,l}+\sigma^2\biggr),
    \label{fp}
\end{multline}
where
\begin{equation}
    \beta_k^{(t)} = \frac{\Re\biggl\{\sqrt{P_{dl,k}}\lvert\boldsymbol{v}_{dl,k}^H\boldsymbol{G}_{k}\boldsymbol{w}_{dl,k}\rvert\biggr\}}{\sum_{l\neq k}\lvert\boldsymbol{v}_{dl,k}^H\boldsymbol{G}_l\boldsymbol{w}_{dl,l}\rvert^2 P_{dl,l}+\sigma^2}.
\end{equation}
As the SINR and the total power constraints are also convex in terms of $P_{dl,k}$ and $P_{comp,k}$, the optimization problem is now convex in terms of $P_{dl,k}$ and $P_{comp,k}$, which can be solved using CVX. Similar to the uplink latency minimization design, by iteratively optimizing $\boldsymbol{W}_{dl,j}$, $\boldsymbol{V}_{dl,j}$, and $\boldsymbol{P}_j$, the minimized computation processing and downlink latency is achieved.

\section{Simulation and results}
\label{sec:simulation and results}
In this section, numerical examples are presented to validate the effectiveness of the proposed system model and algorithms. As shown in Table \ref{table1}, the simulation parameters are carefully designed to reflect real-world conditions, with specific consideration of practical factors such as rainfall attenuation, atmospheric gas attenuation in a harsh environment, and actual turbine positions deployed in the Hornsea project.

\begin{table}[h]
    \centering
    \caption{Simulation Parameters}
    \renewcommand{\arraystretch}{1.1} 
    \small 
    \resizebox{\columnwidth}{!}{
        \begin{tabular}{|p{4.7cm}|p{4cm}|} % Adjust column widths for better fit
            \hline
            \textbf{Parameter} & \textbf{Value} \\ 
            \hline
            Carrier frequency $f_{c}$ & 3.85 GHz \\ 
            \hline
            Total number of waypoints $J$ & 7 \\ 
            \hline
            Downlink Tx power & 40 dBm\\ 
            \hline
            Tx (FBS) antenna height & 1000 m \\ 
            \hline
            Rx (turbines) antenna height & 190 m \\ 
            \hline
           Locations of turbines & See references in \cite{hornsea_planning} \\ 
            \hline
            Tx panel size & 
                $4 \times 4$\\ 
            \hline
          Rx panel size & $3 \times 3$ \\ 
            \hline
            SINR threshold $\text{SINR}_{th}$ & $-10$ dB \\ 
            \hline
             Rainfall Attenuation & 0.026 dB
 \\ 
            \hline
              Atmospheric Gas Attenuation
 & 0.020 dB
 \\ 
            \hline
              Free Space Path Loss
 & 110.643 dB
 \\ 
            \hline
            Receiver Antenna Gain &
            1.761 dB
\\
\hline
Noise power & $-170$ dBm\\
\hline
UAV type & Airbus Helicopters VSR700
\\
\hline
Cruse speed & 220 km/h \\
\hline
Flying distance from the first to last waypoint & 53.64 km \\
\hline
Uplink data size $D_{ul,k}$ \footnotemark[1]& 1 bit\\
\hline
Downlink data size $D_{dl,k}$ & 1 bit\\
\hline
Computational intensity $\mu$ & 1\\
\hline
Energy efficiency of the processor $\varsigma$ & 0.8\\
\hline
        \end{tabular}
    }
    \label{table1}
\end{table}
\footnotetext[1]{Both uplink and downlink payloads are assumed to have an uniform size of 1 bit for demonstration. While the actual data size may vary depending on the information exchanged between the turbines and the FBS. This assumption does not affect the feasibility or validity of our proposed algorithm.}

\begin{figure}
    \centering    \includegraphics[width=0.85\linewidth]{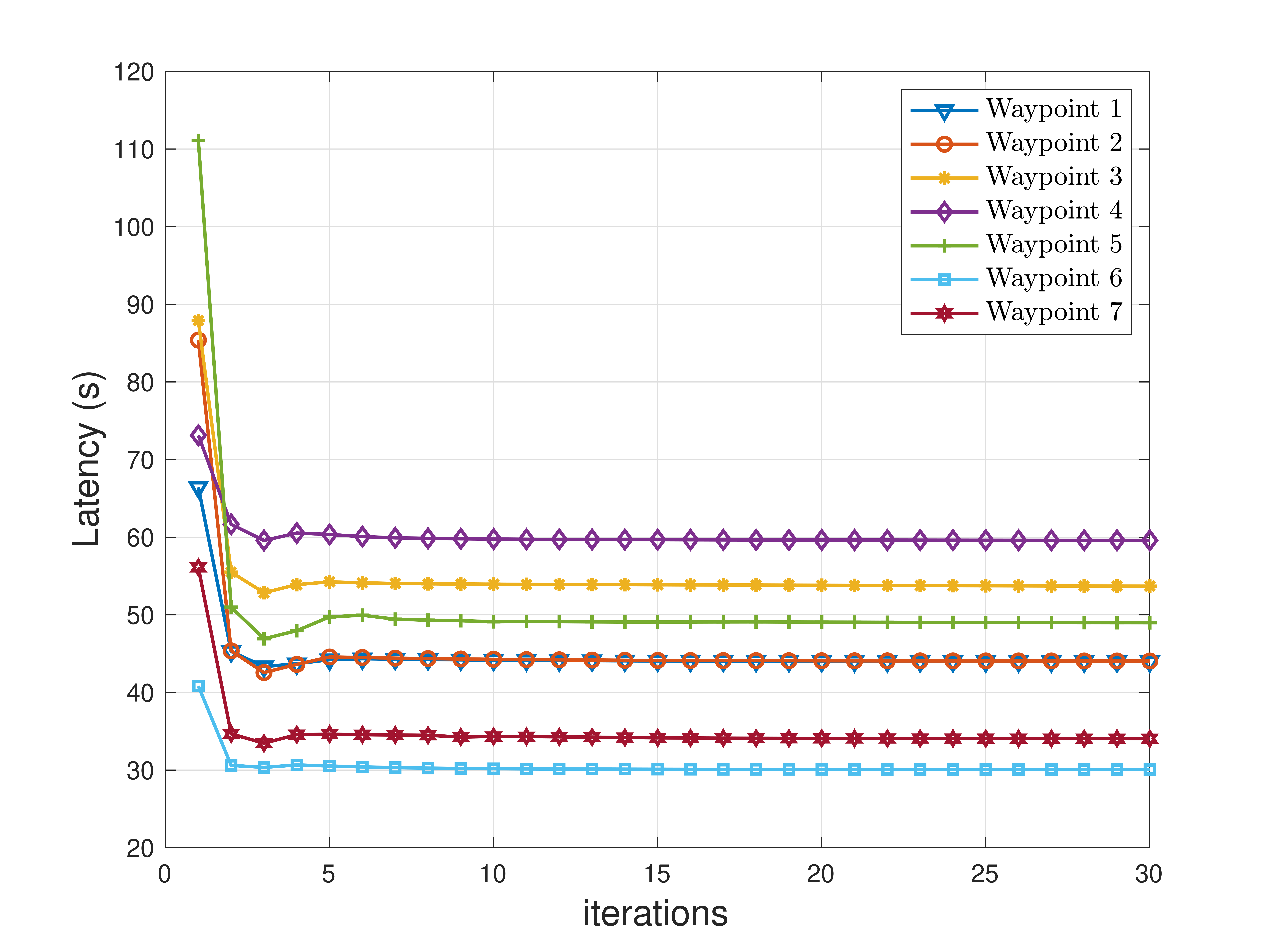}
    \vspace{-3mm}
    \caption{Convergence behaviour at each waypoint.}
    \label{convergence}
    \vspace{-5mm}
\end{figure}

Fig. \ref{convergence} illustrates the convergence behavior of the proposed algorithm, validated at each waypoint by considering both uplink and downlink transmissions. The results indicate that the overall latencies exhibit an initial decline and subsequently stabilize to a consistent value within a finite number of iterations. Variations in latency across waypoints are attributed to differences in the number of turbines and channel conditions at each location.

\begin{figure*}[t]
\centering
\begin{minipage}{.32\textwidth}{
\includegraphics[width=2.4in,height=1.8in]{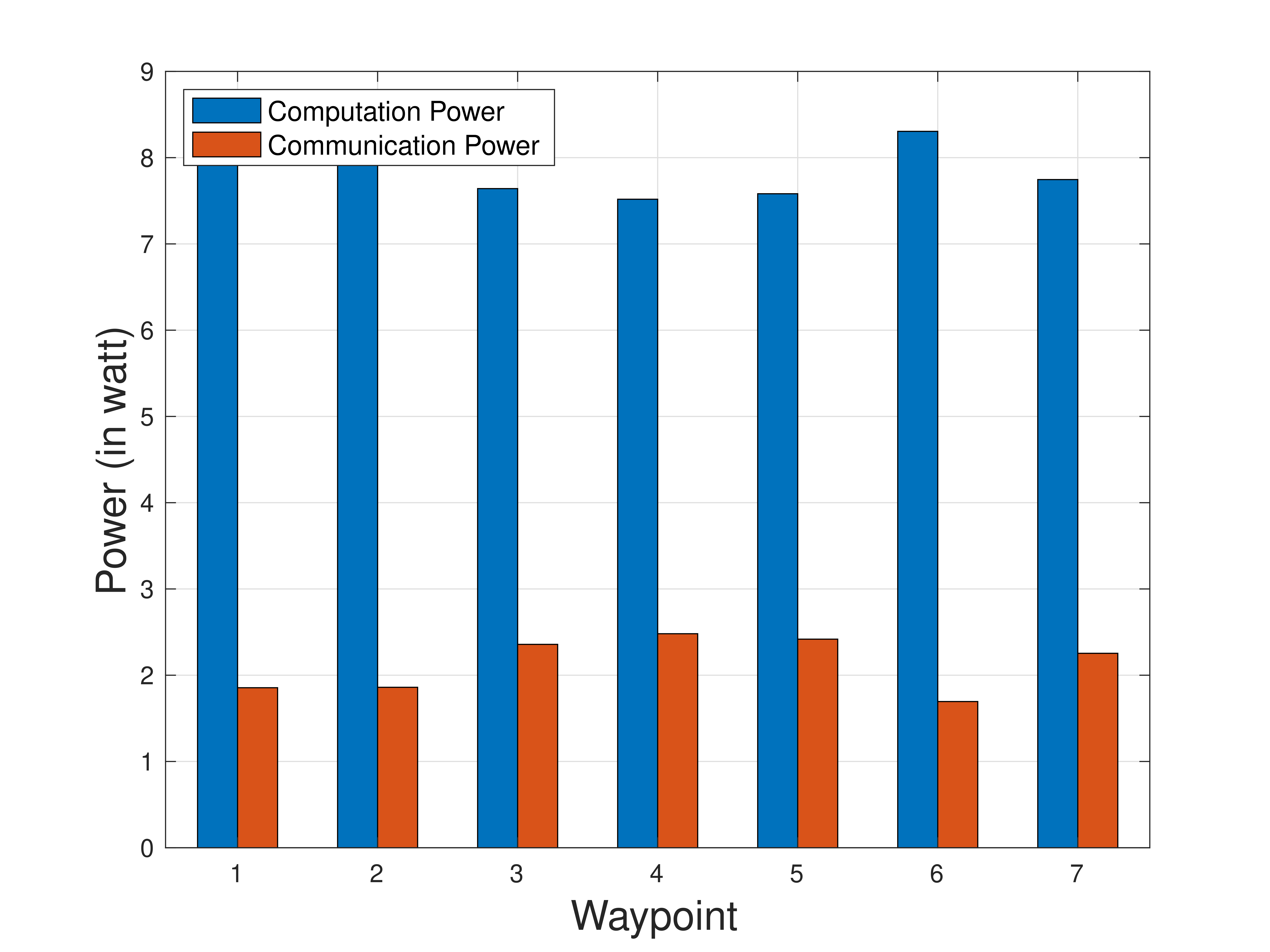}}
    \caption{Power allocation at each waypoint.}
    \vspace{-5mm}
    \label{power_alloc}
\end{minipage}
\begin{minipage}{.32\textwidth}
{\includegraphics[width=2.4in,height=1.8in]{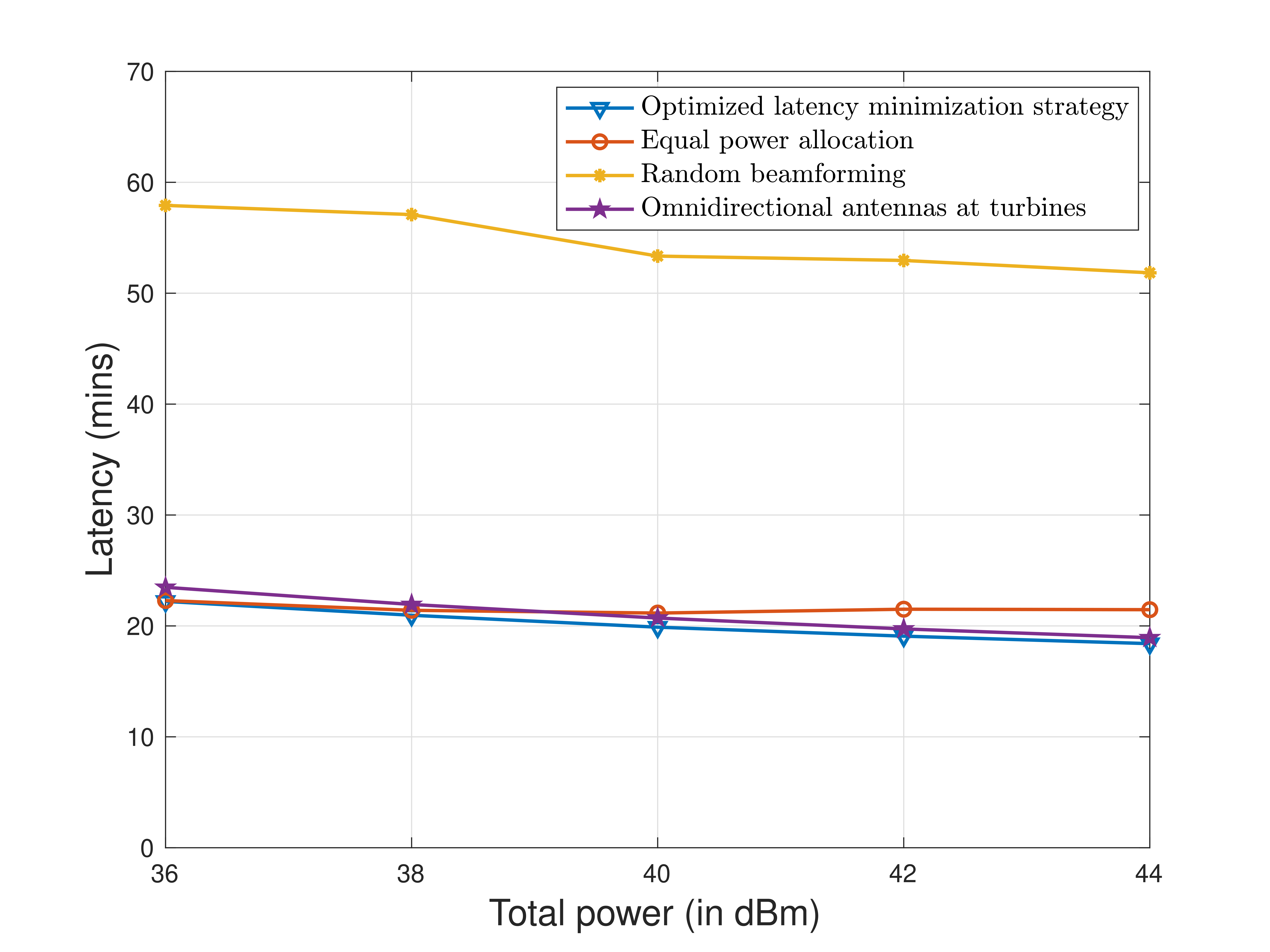}}
    \caption{Comparison of the realized latency across different strategies under different power.}
    \label{power_plot1}
    \vspace{-5mm}
\end{minipage}
\begin{minipage}{.32\textwidth}
{\includegraphics[width=2.4in,height=1.8in]{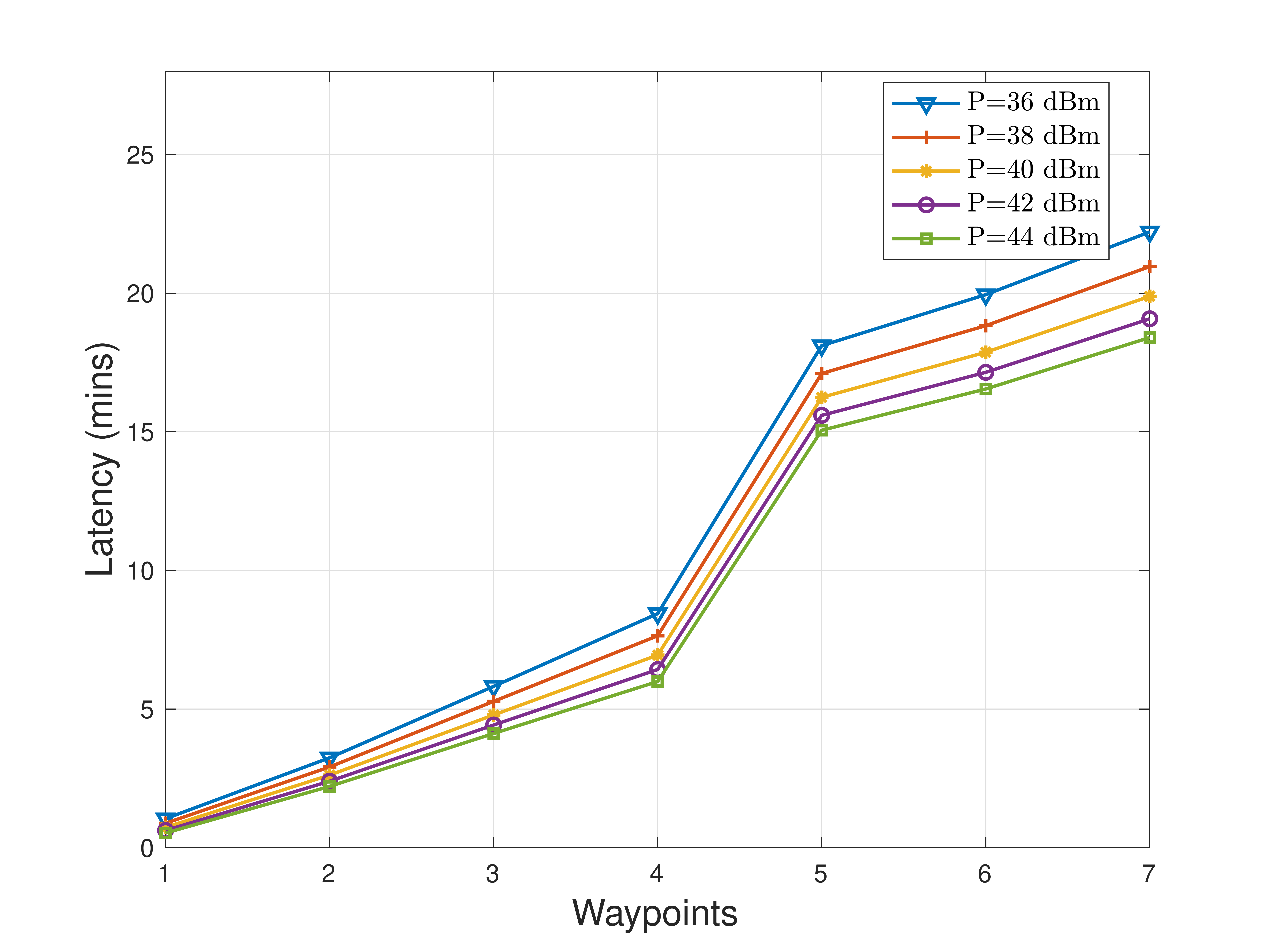}}
    \caption{Cumulative latency at each waypoint.}
    \label{power_plot2}
    \vspace{-5mm}
\end{minipage}
\end{figure*}

In Fig. \ref{power_alloc}, we study the power allocation of the FBS at each waypoint during the downlink transmission. Fig. \ref{power_alloc} depicts that the power is adaptively allocated to computation and communication tasks in order to realize the minimized latency. The majority of the power is allocated to computation, suggesting that the system prioritizes processing over transmission, while the SINR constraint ensures the communication quality. 

Fig. \ref{power_plot1} examines the impact of the total transmit power at the FBS. The proposed design is compared with three baseline scenarios: (a) equal power allocation between computation and downlink transmission tasks, (b) random uplink and downlink beamforming design, and (c) omnidirectional beamforming at each turbine. The results indicate that our proposed design consistently yields the lowest latency across all power levels. In contrast, random beamforming design can lead to excessively high latencies, exceeding 50 minutes. Furthermore, optimizing power allocation achieves lower latency compared to equal power allocation, especially when the power increases. The omnidirectional beamforming at each turbine strategy, which is the general case in practical implementation, achieves the second-best performance in most cases. The latency achieved with omnidirectional antennas is very close to that of the proposed design, particularly at higher power levels.

Fig. \ref{power_plot2} illustrates the relationship between the transmit power at FBS and the cumulative latency at various waypoints. As the transmit power increases from 36 dBm to 44 dBm at each waypoint, a noticeable reduction in latency is observed across all waypoints. This trend suggests that higher transmit power enhances the efficiency of both computation and communication, leading to lower latencies. It can be seen that the highest latencies occurred at lower power levels (36 dBm and 38 dBm) and the lowest latencies occurred at higher power levels (42 dBm and 44 dBm). 

\section{Conclusions}
\label{sec:conclusion}
In this paper, we presented a comprehensive framework for FBS to enhance real-time monitoring and control of offshore wind farms, specifically focusing on the UK Hornsea offshore wind farm project. Given the challenges associated with large-scale offshore deployments, such as long-distance communication, harsh weather conditions, and the lack of permanent infrastructure, we propose the FBS-assisted system, which provides a mobile, flexible, and cost-effective solution via resource optimization. Simulation results demonstrate that a 5G FBS solution can gather monitoring data and control information in approximately 20 minutes for a wind farm spanning 53.64 km with 173 turbines. The developed latency-aware, resource-optimized approach improves the overall efficiency, reliability, and responsiveness of wind farm monitoring and control.

\bibliographystyle{IEEEtran} %

\bibliography{IEEEabrv,references} 

\end{document}